# A VM-Agnostic and Backwards Compatible Protected Modifier for Dynamically-Typed Languages

Iona Thomas[a], Vincent Aranega[a], Stéphane Ducasse[a], Guillermo Polito[a], and Pablo Tesone[a]

a  Univ. Lille, Inria, CNRS, Centrale Lille, UMR 9189 – CRIStAL, France

**Abstract**
In object-oriented languages, method visibility modifiers hold a key role in separating internal methods from the public API. Protected visibility modifiers offer a way to hide methods from external objects while authorizing internal use and overriding in subclasses. While present in main statically-typed languages, visibility modifiers are not as common or mature in dynamically-typed languages.

In this article, we present PROTDYN, a self-send-based visibility model calculated at compile time for dynamically-typed languages relying on name-mangling and syntactic differentiation of self vs non self sends. We present #Pharo, a PROTDYN implementation of this model that is backwards compatible with existing programs, and its port to Python. Using these implementations we study the performance impact of PROTDYN on the method lookup, in the presence of global lookup caches and polymorphic inline caches. We show that our name mangling and double method registration technique has a very low impact on performance and keeps the benefits from the global lookup cache and polymorphic inline cache. We also show that the memory overhead on a real use case is between 2 % and 13 % in the worst-case scenario.

Protected modifier semantics enforces encapsulation such as private but allow developers to still extend the class in subclasses. PROTDYN offers a VM-agnostic and backwards-compatible design to introduce protected semantics in dynamically-typed languages.

## The Art, Science, and Engineering of Programming

# 1 Introduction

Method visibility modifiers such as public, protected, and private are present in mainstream statically-typed object-oriented languages such as Java, C#, and C++. These modifiers specify if and how a method should be used in other parts of an application. It is well-known that some methods may be defined for internal usage only and therefore should be differentiated from the ones that are available publicly [33, 25, 34, 28]. Visibility modifiers also define whether a method can be overridden.

In dynamically-typed object-oriented languages, visibility modifiers are not as common and not as mature. Python implements private methods through name-mangling and does not enforce protected methods [27]. Ruby dynamically enforced access modifiers have changed their semantics recently in 2019 [29]. An ECMAScript proposal in stage 4 as of the writing of this paper proposes the inclusion of private fields [13]. All methods are public in Smalltalk and several of its descendant [18, 36, 5].

Protected method modifiers in their general form offer two interesting facets: (1) they hide methods from external objects while allowing the class and its subclasses to invoke protected methods, and (2) they allow redefinition in subclasses to support reuse [34]. This dual aspect makes them an interesting concept that fits well with late-bound object-oriented languages. This raises the question of their introduction in object-oriented dynamically-typed languages (*i.e.,* without static type checking).

In this paper, we present PROTDYN, a self-send-based visibility model that introduces protected methods in dynamically-typed languages and computes method visibility at compile time. That is, we distinguish syntactically between self sends (self doSomething) and object sends ( (anObject doSomething) following Schärli *et al.* analysis [31, 37]. Such syntactic differentiation help to statically predict the message receiver's type and determine whether invoking protected methods is legal or not. Object-sends can only invoke public methods, while self-sends can invoke protected and public methods. In addition, subclasses can make method access more permissive, by changing a modifier from protected to public, but not the opposite. This makes our visibility solution determinable at compile-time based on syntactic information.

We present an implementation of PROTDYN in Pharo named #PHARO. This implementation is (1) optionally loadable, (2) does not require any changes to the default method lookup supported by a virtual machine, and (3) is backwards compatible with existing code.

We show that our implementation based on name mangling and double method registration introduces a marginally noticeable overhead of around 1 % compared to using Pharo without our modifier. We analyse the behaviour of #PHARO in the presence of optimizations such as global lookup caches and polymorphic inline caches, and its memory usage. Our results show that our implementation is practical (Section 5 and 6). Appendix E presents a Python implementation.

**The contributions of this article are:**

- The definition of a protected method model named PROTDYN;

Iona Thomas, Vincent Aranega, Stéphane Ducasse, Guillermo Polito, and Pablo Tesone

- #Pharo, an implementation for Pharo [5] (a Smalltalk descendant) that supports optional protected modifiers, relies on default method lookup with negligible run-time overhead, and that applies to other dynamic object-oriented languages;
- evidence that a name mangling solution with double registration produces negligible performance overhead: it profits from lookup optimisations such as global lookup caches and polymorphic inline caches;
- evidence that a name mangling solution with double registration produces a memory usage overhead that is practical, with up to 13 % in worst case measured on a case study.

**Outline** Section 2 presents the state of protected modifiers in a selection of object-oriented languages and presents our model. Section 3 describes the core elements of our model ProtDyn. Section 4 describes #Pharo, the implementation of this model in Pharo. Section 5 presents a run-time performance analysis of our solution and Section 6 studies its use of memory. Section 7 discusses alternative implementations and Section 8 compares to related work, before concluding in Section 9.

## 2 About Protected Methods

The protected modifier supports encapsulation [31, 30] while still allowing overriding [9, 34]. Although the protected method modifier is present in many mainstream object-oriented languages, their semantics are slightly different in each of them.

### 2.1 Illustrating Existing Protection Modifiers

While the protected modifier generally conveys that a method is hidden from clients and accessible from subclasses, several variations exist in mainstream languages. Programming languages propose different visibility modifiers that have similar names yet they have different visibility semantics, redefinition semantics, and enforcing mechanisms. Notably:

- In Java, a protected method is visible from its class, its subclasses and classes within the same package. They are overridable by subclasses. Their access is statically enforced by the type system [38].
- C++'s protected modifier is similar to Java's, except for visibility by friend declaration that replaces package visibility [23].
- In Ruby, PHP, and C# protected methods have neither package visibility nor friend mechanism. In C# their access is statically enforced by the type system. While in Ruby and PHP, they are dynamically enforced.

In Section 8.1 we provide a larger and deeper comparison of modifier semantics.

## 2.2 Protected Semantics for Dynamically-Typed Languages

In this paper, we propose a self-send-based visibility semantics with a compile-time implementation mechanism for dynamically-typed languages based on the syntactic differentiation between self-message sends and object-message sends. Note that Appendix A contains definitions of terms we use in the rest of this paper. Our design choice to distinguish between self-message sends and object-message sends is guided by the analysis of Schärli *et al.* [30] that we summarize below.

### 2.2.1 Visibility Semantics by Schärli et al.

Schärli *et al.* [30] studied different semantics for a modifier's visibility: based on static types, class-based, identity-based, and self-send based. Schärli's analysis considers only locally-bound private methods, but the different syntactic options are still relevant for protected methods.

**Static type.** The first option relies on a static type system. In terms of scope, a visibility modifier can be implemented to restrict access to instances of a single class. However, the static type requirement is opposed to the nature of the dynamically-typed languages that we are targeting.

**Dynamic Class-based.** The second option is a class-based modifier: the visibility of a method is assessed at run time according to the class of the sender object and the class where the method has been defined. This can be used to restrict visibility to a single class, or a class hierarchy (the class where the method is defined and its subclasses). With this option, overriding a public method as protected makes reasoning on the program execution more difficult.

**Dynamic Identity-based.** The third option is an identity-based modifier: a sender object can send a protected message *only to itself*. The receiver's identity of a protected method is compared at run time with the sender's identity. Such semantics bring subtle differences in method executions that may make programs difficult to predict. It can lead to a program working with some instance of a class and failing with other instances of the same class. In addition, using the same selector on the same object can produce a different result if a protected method is accessed in one case, and a public method for the same selector in another case.

**Self-send-based.** The final option is self-send-based visibility: protected messages can only be sent to self. As this does not rely on dynamic information on the receiver, it allows one to decide at compile time which message-send sites can access methods with restricted visibility. Self-send-based visibility makes it easier for the programmer to understand quickly where protected methods can be accessed and where they can not. The lookup is It makes reasoning on code easier.

Schärli et. al still points out some issues with symmetry properties. For example, self = arg might not be equivalent with arg = self even if arg and self point to the same object. This happens for the same reason as the problems with the two previous visibility semantics, but at least, in this case, the developer can identify syntactically whether a protected method will be invoked.

**2.2.2 Chosen Semantics for a Protected Modifier**

We want a protected modifier with the following properties:

- A **protected** method is visible from a message-send site if the receiver is the same as the current method receiver (being of the same class is not enough).
- A **protected** message-send site is statically determined using syntactic information.
- A **protected** method is overridable from the subclasses of the defining class.

We chose a self-send-based visibility semantics as it creates fewer ambiguities than the identity-based one. We say that the semantics above are self-send-based because they do not take into account lexical scoping such as the class, package, or namespace as a design choice and they only rely on the fact that the receiver is syntactic self/super or not. We also add the following rule:

- A **public** method can not be overridden by a **protected** method in any subclass.

This rule supports subtyping between a class and its subclasses. In addition, it ensures that the symmetry issues will not arise with self = arg and arg = self as long as arg and self point to the same object.

**ProtDyn in a nutshell.** ProtDyn is a self-send-based time visibility model determined at compile time where:

- A **protected** method is visible from a self-send site.
- A **public** method (*i.e.,* non-protected) is visible from all sites (object-send and self-send sites).
- A **protected** method is overridable from the subclasses of the defining class.

Notice that this model trades off programmers' flexibility for a compile-time mechanism based on syntactic information. Protected methods are never visible from a non-self send (`anObject doSomething`), even though the receiver object may be identical to the current receiver (`anObject == self`).

# 3 ProtDyn: A Protected Modifier Model

This section introduces an informal description of ProtDyn, our protected modifier model based on the syntactical differentiation of object-sends and self-sends [30]. Appendix D presents a formal definition of this model.

## 3.1 Object-Send and Self-Send Lookup Semantics by Example

ProtDyn differentiates send sites by their syntactic receiver, thus object-sends and self-send behave differently:

- Object-sends see only *public* methods.
- Self-sends see *public and protected* methods.

Figure 1 illustrates, with six simple scenarios, the distinction between object-sends and self-sends. In those scenarios, aA represents an instance of the A class, and aB, an instance of the B class. We also denote with A»foo the method with selector foo in class A. Additional explanations are available in Appendix B.

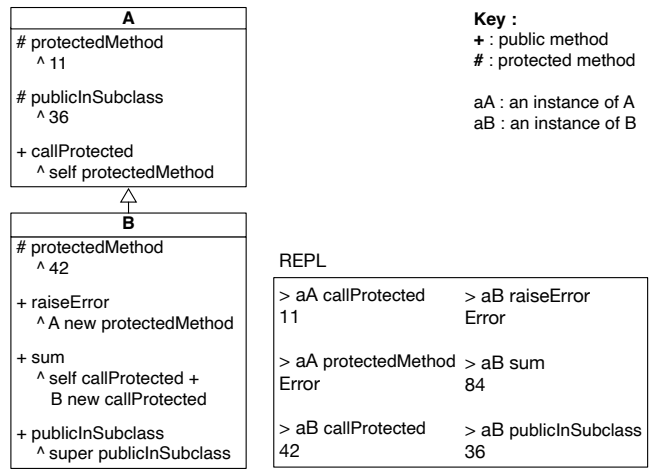

**Figure 1** Message sending is modified to distinguish between object-sends and self-sends: only object-sends can invoke protected methods which can also be overridden and taken into account by default method lookup.

This model trades off programmer's flexibility for a compile-time-based visibility mechanism. For example, in a statement sequence such as temp := self. temp foo, the foo send will have more restricted visibility than directly doing self foo. Syntactic differentiation is a simple rule that can be easily learned by developers and makes program understanding easier [30].

### 3.2 Changing Visibility in Subclasses

Reducing the visibility of a method in a subclass makes programs difficult to reason about. Figure 2 shows two different examples where reducing visibility creates ambiguities when the method lookup in object-sends skips protected methods. In Figure 2(a), aB sum returns 108. The self size self-send lookup finds the B»size method as expected, but the object-send B new size cannot find the B»size method as it is protected. The lookup finds the A»size method which is public. In Figure 2(b), reducing the visibility of the sum: method makes the result of a sum: send asymmetrical:

- aB sum: aA returns 77 as expected. The self-send finds the protected B»size method and the object-send lookup finds the A»size method.
- aA sum: aB returns 132. The self-send lookup finds the A»size method as expected, but the object-send cannot find the B»size method as it is protected. Therefore, the lookup finds the A»size method a second time.

There are two different alternatives to avoid this unintuitive behaviour, left as an implementation choice for language designers:

Iona Thomas, Vincent Aranega, Stéphane Ducasse, Guillermo Polito, and Pablo Tesone

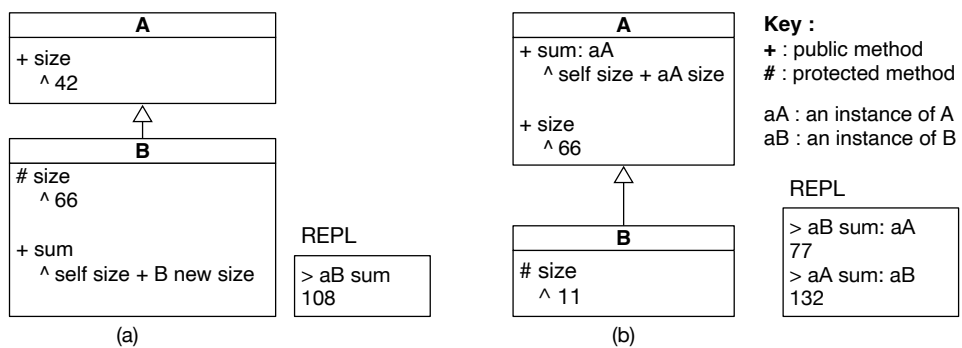

**Figure 2** Overriding a public method by a protected one leads to buggy situations where self-sends and object-sends can yield two different results. (a) superclass method is used, (b) asymmetrical results.

- *Raising an error at run time.* For object-sends, the semantics of this solution is that the lookup throws an exception if it finds a protected method. This is the strategy chosen by the Ruby language.
- *Statically forbidding the visibility change in subclasses.* This solution avoids the problem by construction and is used by statically-typed languages such as Java.

The other way around, opening up the API by making a protected method public in a subclass does not lead to the discussed issues because public methods were visible from self-sends too. This provides additional flexibility by giving developers a better way to expose previously protected methods. This ability is present for example in the Java language.

## 4 Implementation

In this section, we present #Pharo, an implementation of ProtDyn for the Pharo programming language. We first present the design principles behind our implementation and an overview of our solution. We then delve into the two key aspects of our implementation: double public method registration and self-send selector mangling. Moreover, as a demonstration of the applicability of ProtDyn to other dynamic languages than Pharo, we ported our solution to Python (See Appendix E for more details).

### 4.1 Design Principles

The design behind our protected modifier implementation follows the principles:

- **Backward compatible:** *Existing* programs (*i.e.,* not using protected modifiers) should continue to work and expose the same behaviour under the presence and absence of protected method support.

# A VM-Agnostic and Backwards Compatible Protected Modifier f. Dynamically-Typed Lang.

- **Not requiring a new runtime:** The solution should not be based on changing the virtual machine execution logic, to ease portability and deployment.
- **No run-time penalty when...**
  - **not using protected methods:** A program not using protected modifiers should not be impacted by the presence of the protected modifier implementation.
  - **using protected methods:** A program using protected methods should not have a run-time penalty compared to using public methods only.

## 4.2 Implementation Overview

■ **Listing 1** Code excerpt from Figure 1.

```
1  B » protectedMethod [
2    <protected>
3    ^ 42 ]
4  B » raiseError [
5    ^ A new protectedMethod ]
6  B » sum [
7    ^ self callProtected + B new
       ↪ callProtected ]
8  B » publicInSubclass [
9    ^ super publicInSubclass ]
```

Our implementation is based on two techniques: *double registration of public methods* and *selector mangling of self-sends*. Protected methods are identified by developers using method annotations using the Pharo annotation system [11], illustrated in Figure 1.

Protected methods are registered in the method dictionary of their class with a mangled selector. Public methods are registered two times, once with their original selector and once with their mangled selector. All self-send sites are rewritten by mangling the message selector. These two transformations are shown in Figure 3. Section 4.5 presents how to limit the recompilation to classes using protected methods and their subclasses.

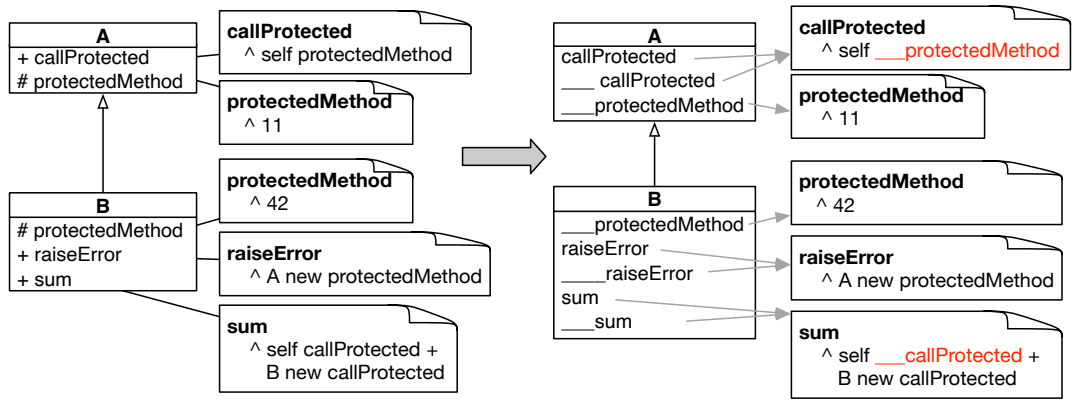

■ **Figure 3** From the model to the actual implementation: Selector mangling and public method double addition to their class method dictionary.

## 4.3 Double Public Registration

Protected methods are added to method dictionaries with a mangled selector (*e.g.,* prefix __ in the following examples and Figure 3). To avoid conflicts, mangled selectors are forbidden by the compiler and are not meant to be used directly. Self-sends are

Iona Thomas, Vincent Aranega, Stéphane Ducasse, Guillermo Polito, and Pablo Tesone

modified at compile time to look for methods with this prefix (see Section 4.4). Protected methods are therefore visible from self-sends, but not from object-sends.

In addition, public methods are visible from both self-sends and object-sends (*i.e.,* all sends). Self-sends see the methods installed with the mangled selector while object-sends see methods installed with their original selector. As illustrated in Figure 3, in class A:

- The protectedMethod appears only once in the method dictionary : at the __protectedMethod selector.
- The public method callProtected appears twice : at callProtected without prefix and __callProtected with the prefix.

### 4.4 Selector Mangling for Self-Sends Sites

During compilation, all self-sends are mangled. In Appendix Section 4.4, we formalize this transformation. Here are key points related to selector mangling for self-sends:

- Self-send sites are rewritten to use mangled selectors to see protected methods. For example, method A»callProtected contains a self-send to protectedMethod, that is rewritten as a send to __protectedMethod. The same goes for super-sends.
- Mangled self-sends see public methods because they are installed with the mangled selector in addition to the original selector. For example, B»sum definition contains a self-send to callProtected, hence it is rewritten as a send to __callProtected. The same goes for super-sends.

The rewrite unit of our implementation is a class hierarchy. If a class is using the protected modifier for the first time, we chose to recompile all methods in the class and its subclasses and add all the duplicated mangled entries when we add a protected method for the first time. This avoids recalculating the exact set of methods to be recompiled for each new protected method which would be required for lazy recompilation.

### 4.5 Preventing Selector Mangling Propagation to the Whole System

Our implementation rewrites all self-sends in a whole *descending* hierarchy to avoid unnecessary rewrites of the existing system. Superclasses without protected methods do not have double registration of public methods, thus self-sends will not find mangled entries in their method dictionaries. Mangling selectors upwards in the hierarchy would propagate the transformation to the full system.

To avoid this problem, our rewriting strategy performs an additional check at compile-time when mangling self-sends. We handle public methods defined above classes with protected methods differently from other public methods. This is the case of the method copy in Figure 4: it is publicly defined in Object and used as a self-send in class A. For such a method, we do mangle the self-send in the classes defining protected methods. In Figure 4, the method protectedMethod invokes copy without mangling as we do otherwise (and explained above). This way we do not have to recompile Object and therefore avoid the propagation of the double registration to the entire system. Applying double registration only on classes using the protected modifier limits the memory overhead (See Section 6).

As the modifier of a method can only change from protected to public in subclasses, any redefinition of copy will be public (as in class B). This ensures that invoking protectedMethod on an instance of B correctly finds the redefined version of copy in B.

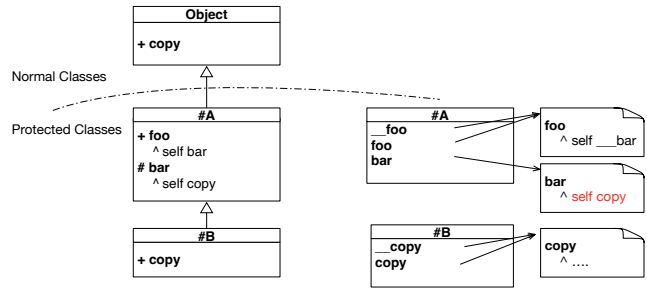

**Figure 4** Limiting the propagation of recompilation with selector mangling for protected to the top of the hierarchy.

A special case appears when a subclass does a self-send of a non-existing method such as the message self unknown in method anyMethod of class A in Figure 5. In that case, our implementation assumes it is a public send and the message-send site will be recompiled when (if) that method is installed later.

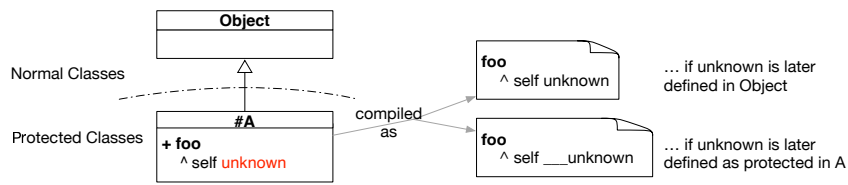

**Figure 5** Compiling a method with an undefined selector assumes a public message.

## 5 Performance Evaluation

We evaluate our solution both in terms of its impact on performance and memory usage. This section evaluates the performance impact, while the following section does it on memory usage (See Section 6).

### 5.1 Experimental Design and Methodology for Speed Performance

Informally speaking, the main performance impact of our solution would come from an increase in the number of entries in method dictionaries, which would introduce a negative impact on CPU caches, method lookup algorithms, and message sends. The goal of this evaluation is to assess such an impact. In our evaluation, we compare several benchmarks on three different scenarios using three different virtual machine configurations derived from the Pharo VM v8.1.0-alpha-335-g70b7e3542.

**Scenario 1: Method lookup impact**   Using our implementation, public methods increase the size of the selector namespace by installing each method once with the non-protected and once with the protected selector. In the *worst case* where all methods are public, they double the method dictionary number of keys. This is the case we measure in all performance benchmarks. Thus, we measure the performance impact on lookup in a token-threaded interpreter implementation of Pharo [14], with lookup caches disabled [10].

**Scenario 2: Lookup cache impact**   Global lookup caches cache method lookup results avoiding the subsequent lookup of the same ($receiver type, selector$) pair [10]. This scenario uses the same token-threaded interpreter above with a global lookup cache enabled. The global lookup cache is a hash table with 1024 entries and performs up to three lookups in the cache per message-send before doing a slow method lookup. This scenario is two-fold: we compare the impact on run-time performance and cache behaviour (*i.e.,* # hits and misses).

**Scenario 3: Polymorphic inline cache impact**   Polymorphic inline caches further avoid lookups by localizing a lookup cache on each message send-site, typically implemented with machine code stubs and code patching [21]. This scenario uses a mixed-mode Pharo implementation combining a token-threaded interpreter, a 1024-entry global lookup cache, and a non-optimizing method JIT compiler with polymorphic inline caches.

All scenarios are run in three different configurations:

- **Baseline** #Pharo not installed (and thus not used).
- **Case 1** #Pharo is installed but not used.
- **Case 2** #Pharo is installed and used in a *worst-case scenario*. As we want to estimate of the maximum run-time penalty using #Pharo, all our benchmarks enable #Pharo but let all methods as public. This forces a *double registration for all methods*

### 5.2 Methodology and Setup

We designed our benchmarks following the guidelines of Georges *et al.* [17]. We did our best to minimize the system's noise [3], close all non-related non-essential applications and services and shut down the internet connection. The machine was plugged in and there was no user interaction until the benchmark finished. We ran our benchmarks for #Pharo on a MacBook Pro 17.1 with an Apple M1 processor (8 cores including 4 performance cores and 4 efficiency cores) and 16GB of LPDDR4 RAM, running on macOS 12.0.1.

We use a fixed number of in-process iterations to determine steady-state instead of dynamically detecting it because none of our VM configurations includes profile-guided optimizations that require a long warmup time. Our methodology is as follows:

- 200 VM invocations.
- 55 benchmark iterations for each VM invocation.

- the first 5 benchmark iterations of each VM invocation are discarded (done to warm up the caches), leaving 50 measures per invocation.
- Package loading is done beforehand and saved in a snapshot.

For each benchmark, we report the average run times of each VM invocation. Figures plot numbers relative to the baseline configuration instead of absolute numbers for readability. For each benchmark, we show the distributions with violin graphs complemented with a whisker boxplot showing the median, lower, and upper quartiles. The upper whisker is the minimum between the max relative run time value and the value of the $upper\_quartile + 1.5 \times interquartile\_range$. The lower whisker is the minimum between the min relative runtime value and the value of the $lower\_quartile - 1.5 \times interquartile\_range$. Assuming a Gaussian distribution, 99 % of the values should be inside the whiskers. Any data points outside the whiskers are shown by black dots.

The average relative run times are shown by red dots within each violin shape. This is the main performance indicator used in the analyses.

### 5.3 Selected Benchmarks

We selected four different benchmarks, avoiding on-principle microbenchmarks because they would not exercise message sends and our protected implementation:

**Microdown**   Microdown is an implementation of a Markdown superset defining a parser, document tree and several exporters. The parser uses a delegation-based approach using many polymorphic calls. We wrote a benchmark implementation that parses the `README.md` file of the Pharo GitHub repository.

**Delta Blue**   The DeltaBlue one-way constraint solver. We used the implementation available in the SMark benchmark library [32].

**Richards**   An OS kernel simulation originally written in BCPL by Martin Richards. We used the implementation available in the SMark benchmark library [32].

**Bytecode Compiler Benchmark**   The Pharo bytecode compiler exercises various compilation aspects: parsing, AST generation, semantic analysis, linear IR generation, and bytecode generation. During the benchmark, we perform several method compilations with source code larger than typical Pharo methods. We used the benchmark implementation available in the SMark benchmark library [32]. Notice that this benchmark exhibits different behaviour than the previous three since our protected method implementation introduces modifications in the compiler.

Full results are available in Appendix F.

### 5.4 Scenario 1: Lookup Performance

We evaluate the impact of #Pharo on lookup performance using the VM with a disabled global lookup cache. This means that each message-send instruction produces a lookup in the class hierarchy happens. Figure 6 shows the distribution of the relative run times of three benchmarks (Microdown, Deltablue, and Richards). Figure 7 shows the performance of the Compiler benchmark.

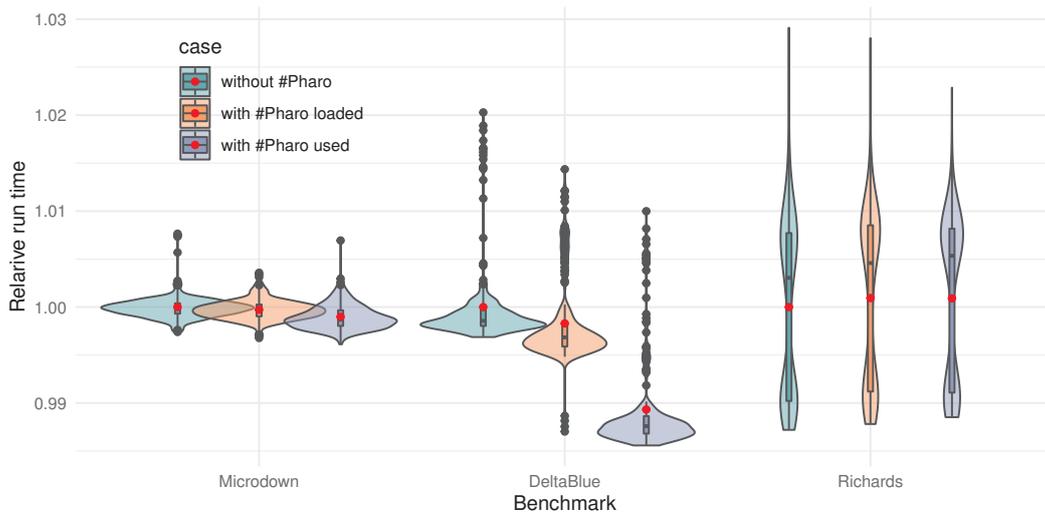

**Figure 6** Relative run time performances with global cache disabled on Microdown, Deltablue and Richards benchmarks. Lower is better. A red dot marks the average.

For Microdown and Richards benchmarks, the average relative variation is less than 0.1 % when #Pharo is loaded and when it is used in the worst-case scenario. For the DeltaBlue benchmark, we have an average run time with #Pharo used that is at 98.9 % of the average run time without #Pharo. The run-time distribution is consistent for each benchmark.

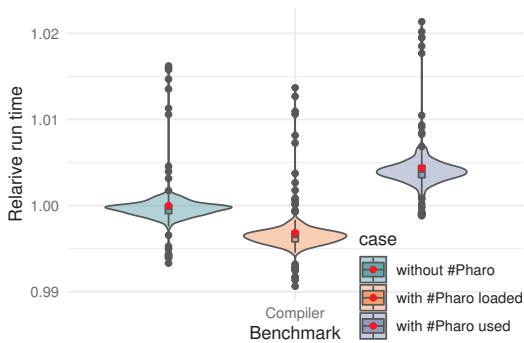

**Figure 7** Relative run time performances on the VM without global cache for the Smark Compiler benchmark. Lower is better. The red dot marks the average.

The compiler benchmark has been set aside because our approach is implemented as a compiler plugin, introducing variations between #Pharo loaded and unloaded. When #Pharo is loaded the average impact at compile time is a 0.3 % speed-up.

When the compiler package itself uses #Pharo, we observe a 0.4 % slowdown. This is the biggest slowdown compared to the other benchmarks on the VM without a global cache.

We observe small differences in the run time performance for which we do not have an explanation yet. When these small performance changes were significant, their impact was usually below 0.5 % of the average speed without #Pharo and always below 1.1 %. *This shows that our*

# A VM-Agnostic and Backwards Compatible Protected Modifier f. Dynamically-Typed Lang.

*prototype is a viable solution and that we do not introduce significant run-time costs related to the lookup.*

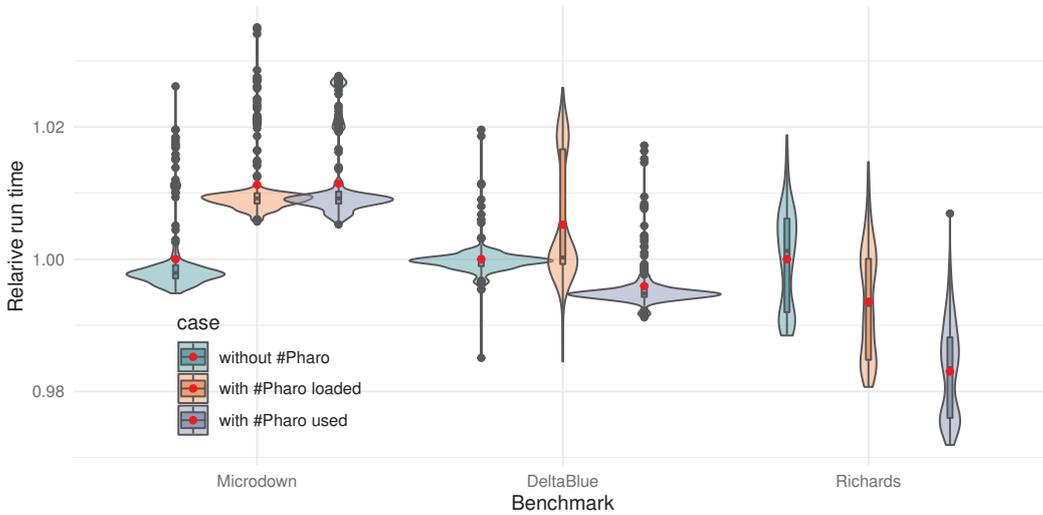

**Figure 8** Relative run time performances on the VM with global cache only for Microdown, Smark Deltablue and Smark Richards benchmarks. Lower is better. A red dot is the average.

## 5.5 Scenario 2: Global Lookup Cache Performance

This experiment evaluates the impact of the global lookup cache on performance. We run the same benchmarks as Experiment 1, enabling the global lookup cache.

Our results show that the benchmark that is slowed down the most is Microdown (see Figure 8). With #Pharo only loaded and #Pharo used, there is a 1.1 % slowdown in the average run time compared to #Pharo unloaded. Delta Blue with #Pharo loaded is also slowed down by 0.5 %. However, Delta Blue with #Pharo used is sped up by 0.4 %. There is also a speedup on Richards, both with #Pharo loaded (0.6 %) and used (1.7 %).

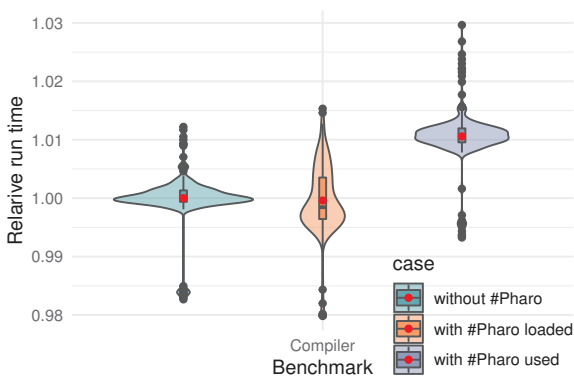

**Figure 9** Results with lookup cache enabled on the Compiler benchmark. Lower is better. Red dots mark the average.

Regarding the compiler Benchmark presented in Figure 9, #Pharo loaded does not introduce slowdowns as we observed in Experiment 1. The speedup using #Pharo increases, from 0.3 % to over 1 % with the global lookup cache.

These changes in relative runtime performances can be related to the number of hits from the global lookup cache. See Section 5.6, where we study the percentage of cache hits and misses.

In summary, in this experiment, we observe slowdowns of at most 1.1 % with a global lookup cache. *This confirms that our prototype performs well in the presence of a global lookup cache*.

### 5.6 Scenario 3: Lookup Cache Behaviour

This experiment evaluates the behaviour of our implementation on global lookup caches, given that name mangling introduces a larger set of selectors (with and without prefixes). We run the same benchmarks as Experiment 1, enabling the global lookup cache and recording cache hits and misses. Notice that Pharo's lookup cache works by probing up to three times in the hash table. A lookup in the table is considered a miss if the third probe fails. Figure 10 shows our results as percentages of hits and misses, discerning between the three hit probes. The higher the percentage of hits the faster the program will run because cache misses will trigger a costly lookup. These measurements are done after the five warm-up iterations.

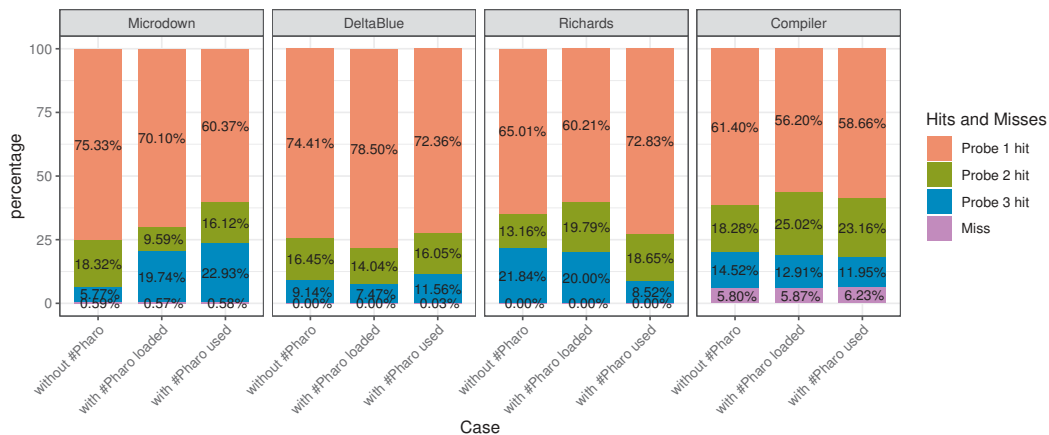

**Figure 10** Cache hits and misses.

In Figure 10, for the Microdown benchmark, we see 5 % fewer first probe hits with #Pharo loaded compared to when it is not, and again 10 % more when it is used. Considering the first and second probe's hits, there are 14 % fewer hits on with #Pharo installed and 3 % more when used. These differences contribute to slowing down the average run time of the Microdown Benchmark seen above.

On the contrary, for the Richards benchmarks with first and second probes, we have 2 % more hits with #Pharo installed and 11.5 % more again with #Pharo used. This explains the observed speedups of the Richards Benchmark shown before.

For the Delta Blue benchmark, we see 4 % more first probe hits with #Pharo installed and 2 % more first and second probe hits. When #Pharo is used we have 2 % fewer first probe hits and 2.5 % fewer first and second probe hits compared to without #Pharo. It matches the run-time variation on this specific VM.

For the Compiler benchmark, we see around 6% misses for all configurations, probably due to hash collisions. There is a slight increase in the percentages in misses: +0.07% with #Pharo installed and +0.4% with #Pharo used compared to without #Pharo. However, considering we doubled the number of selectors in the method dictionary of all the classes of the targeted package, this is a relatively small increase.

Overall there is little to no variation in the number of misses. Performance variations depend on which probe hits: the faster a probe hits, the faster the benchmark will run. Depending on the benchmark, the percentages of probe hits increase or decrease when #Pharo is installed or used. While we cannot extract general rules on when collisions happen while comparing without #Pharo, with #Pharo loaded and with #Pharo used, we can observe that probe hits are related to the changes in run time performances. Therefore, *the variations in run time performances are linked to global cache hits and #Pharo can have a small positive or negative impact on those*.

**5.7 Scenario 4: Polymorphic Inline Cache $PIC$ Performance**

This experiment evaluates the impact of our implementation on Polymorphic Inline Caches (PICs). The informal assumption is that name mangling should not affect the runtime behaviour of PICs. Figure 11 shows the results for the Microdown, Delta Blue, and Richards benchmarks on a VM with JIT and the Polymorphic Inline Cache (PIC) enabled. Figure 12 shows the results for the Compiler benchmark. For the Microdown benchmark, we observe speedups of 0.15% and 0.35% with #Pharo respectively installed and used. The distribution of each case's relative run times is similar. For the Delta Blue benchmark, the distribution of each case's relative run times is also similar, but there are slowdowns of 0.51% and 0.64% with #Pharo respectively installed and used.

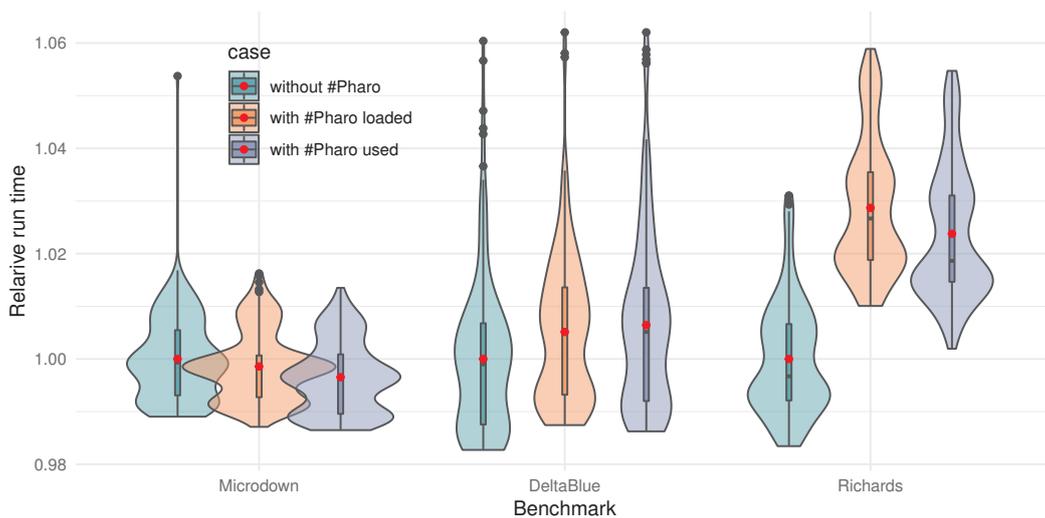

■ **Figure 11** Results on the VM with JIT and PICs enabled for Microdown, Deltablue, and Richards. Lower is better. A red dot is the average.

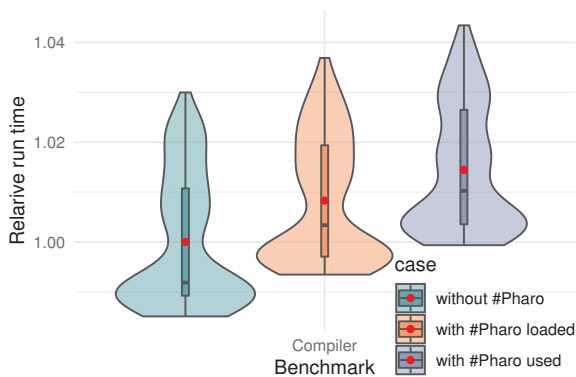

**Figure 12** Results on the default VM for the Compiler benchmark. Lower is better. A red dot is the average.

In the compiler benchmark, we observe an average run time that is respectively 0.8 % and 1.4 % slower compared to the version without #Pharo.

The Richards Benchmark presents the biggest slowdowns, with a run time 2.9 % higher when #Pharo installed and 2.4 % higher when used.

These results reflect the performance of #Pharo on the default Pharo VM with the polymorphic inline cache. In the worst-case scenario, we have an average run time that is less than 3 % longer on the most optimized VM. However, in most cases, average run time variations with #Pharo installed and used are around 1 % or less. As we have seen in previous experiments, those variations can be partially due to the global cache performances and are not directly linked to the installation or the use of #Pharo.

There are variations in average run times, so we cannot claim that there is **no run-time penalty when not using protected** and the **no run-time penalty when using protected** constraint is met with our implementation. However, we believe that the run time performance variations in both cases are low enough to prove that this is a viable implementation.

## 6 Memory Use Analysis of #Pharo

Our solution relies on double registration of methods and name mangling, stressing method dictionaries and selector tables. This section assesses the memory cost of protected Pharo. We use the Microdown application as our benchmarking program (see Section 5.3) and measure the amount of memory used before and after introducing #Pharo and protected methods.

### 6.1 Methodology and Setup

We measure the use of memory using the "Space and time"[35] package. This package traverses the graph of objects starting from Microdown classes and measures the amount of memory taken in the heap in bytes. The traversal stops on global variables to avoid propagating to the whole system. The rest of the setup is the same as the one described in Section 5.1.

We make measurements in three benchmark variations on Microdown which defines 257 classes and 2683 methods.

**#Pharo Unused** Not using #Pharo is used as a comparison baseline.

**Manual Tagging**   We manually tagged 28 methods as protected in 8 classes, belonging to 6 different hierarchies with depths ranging from 1 to 5. Our 28 tagged methods led to 61 classes using our protected modifier, either directly or by inheritance. Few methods could be declared as protected in Microdown because most of its code uses the visitor pattern. Notice that those 28 methods were found by an automatic static analysis of the project. Our static analysis is conservative due to the lack of type information, some potential candidates may have been missed because of unintended polymorphism. Moreover, we made sure that all of the project tests stayed green after tagging: incorrect tagging can overprotect a method and change the application semantics. We leave for future work the analysis of the profitability of protected methods and the automatic migration of applications.

**Worst case**   To get the worst case, all methods are declared as public. This makes our implementation register each method both with mangled and not mangled selectors in all classes, taking more space in method dictionaries and selectors.

## 6.2  Memory Cost: Results

■ **Table 1**  Microdown memory use with and without #Pharo. Absolute numbers in bytes. Results relative to the baseline are in parentheses.

| Measure | Unused | Manual Tagging | Worst Case |
| --- | --- | --- | --- |
| Total size (bytes): | 1 023 216 | 1 044 840 (1.02x) | 1 159 368 (1.13x) |
| Method dictionaries size (bytes): | 249 592 | 263 592 (1.06x) | 326 040 (1.30x) |
| Symbols size (bytes): | 94 920 | 102 528 (1.08x) | 154 560 (1.62x) |
| Number of instances: | 19 563 | 19 835 (1.01x) | 21 321 (1.08x) |
| Number of Symbols: | 3 323 | 3 595 (1.08x) | 5 081 (1.52x) |

Table 1 shows an increase of 2.1 % of the memory used by Microdown in the case where #Pharo is used selectively and 13 % in the worst-case scenario. This increase is due to two main things: first, all the new symbols corresponding to mangled selectors, and second, the size increase of the method dictionaries. By looking at the number of instances, *i.e.,* the number of objects in the studied graph, we can see that all the new instances actually corresponds to the new symbols which are created by name-mangling. As the double registration of the public method creates references to the same compiled method with the selector mangled and not mangled, the number of compiled methods does not increase. We only have a small overhead (48 bytes total in the worst case) in the space taken by compiled methods and compiled blocks. We consider that this is a reasonable increase in memory use that does not threaten the viability of #Pharo.

## 7  Implementation Decisions

This section reports our implementation decisions and some alternatives.

Iona Thomas, Vincent Aranega, Stéphane Ducasse, Guillermo Polito, and Pablo Tesone

## 7.1 Lookup Mechanism Modification

In our implementation, we avoided modifying the method lookup used in the VM by using selector mangling. An alternative implementation is to split the lookup into two different operations, one for public methods and the other for protected methods. To identify the methods we consider the following alternatives:

- Marking a method with its visibility, or;
- Splitting methods into two different collections (*e.g.,* method dictionaries).

Marking methods with their visibility requires the lookup method to iterate all possible methods in a class and differentiate them when using the public or protected lookup. This approach simplifies the structure of the classes but it may affect the performance of the lookup mechanism, as it has to iterate more methods than required. The second alternative, splitting methods into two different method dictionaries, simplifies the lookup mechanism but requires maintaining two different collections per class.

All in all, a modified method lookup algorithm affects all classes in the system, not only the ones using protected methods. Our implementation only affects the classes using protected methods.

## 7.2 Run-Time Visibility Checks

In our implementation, when the lookup for a protected method is performed from an object-send site, the protected method will not be found and an error will be raised. This error is equivalent to the one produced when a message is not implemented, or in Smalltalk terminology, a *Message Not Understood* error.

Alternatively, a more specific error could be raised when a protected method is activated from a object-send site. This is the strategy chosen by Ruby. This alternative requires performing a check on each method activation. We chose to use the lookup solution because it keeps backward compatibility, it profits from existing lookup optimizations, and because *Message Not Understood* errors are already commonly used in Pharo.

# 8  Related Work

Table 2 summarizes the semantics of the protected modifier for some mainstream languages regarding three aspects: their visibility semantics, visibility mechanism, and narrowing. Visibility semantics vary in each language. In what follows, we first focus on Ruby's method visibility modifiers, as it is one of the few dynamic languages offering them. We then discuss Python, which uses name mangling for private modifiers. Then, we present method encapsulation in Javascript and Java, C#, and C++. Finally, we discuss related work on encapsulation in dynamically-typed languages.

■ **Table 2** Accessibility of methods according to visibility modifiers in different languages.

|  | Solution | Visibility | Enforcement | Narrowing |
|---|---|---|---|---|
| static | Java protected | hierarchy, package | compile time | Forbidden |
|  | C++ protected | hierarchy, friends | compile time | Forbidden |
|  | C#protected | hierarchy | compile time | Forbidden |
| dynamically-typed | Smalltalk | all send-sites | N/A | N/A |
|  | Ruby private | hierarchy using self or implicit receiver | run time | Allowed, checked at run time |
|  | Ruby protected | hierarchy | run time | Allowed |
|  | PHP private | class | run time | Forbidden |
|  | PHP protected | hierarchy | run time | Forbidden |
|  | Python private | hierarchy | static (mangling) | N/A |
|  | Python protected | class | convention only | N/A |
|  | JavaScript | class | run-time | N/A |
|  | **ProtDyn** | **hierarchy using self-sends sites** | **static (mangling)** | **Forbidden** |

## 8.1 Existing Visibility Modifiers

**Ruby.** Ruby is one of the few dynamic languages that offers method modifiers: methods can be qualified as public, protected, or private. Ruby syntactically distinguishes private from public methods [29]. Modifiers can be changed in subclasses: a private method may be made public in subclasses opening further the API, or a public method can be restricted to protected or private leading to errors when not called properly. If a subclass uses a stricter modifier, its instances can not be used polymorphically with instances of the superclass.

Ruby's protected methods can be invoked by sending a message from the same class where it has been defined or from its descendants. The receiver can be implicit, self, or another instance from the same family. This means that instances of sibling classes can call protected methods defined in a common ancestor on each other. This is different from our approach and closer to the class-based semantics of Schärli [30].

Ruby's private methods are similar to our protected proposal: in Ruby, private methods can only be called in the class and its subclasses, and only if the receiver is syntactically self or the implicit receiver. In addition, a private method can also be overridden in subclasses. Originally, private methods in Ruby could only be invoked by messages to the *implicit* receiver (*i.e.,* no *self*), restriction removed in Ruby 2.7. Calls to private methods are not statically bound and can be overridden in subclasses.

While the semantics of Ruby's private modifier is similar to #Pharo, the implementation is different: in Ruby, the visibility is checked and enforced dynamically with flags on the methods. A semantic difference with our model is that super is also a valid receiver of protected sends and that we forbid visibility narrowing.

**PHP.** PHP supports also three visibility modifiers which are public, protected, and private. In PHP, private methods can only be invoked from the class where it is defined, but they can be invoked on another instance of the same class. This private semantic is class-based, rather than self-send-based like Ruby's. Note that the official description of the language semantics is rather vague. We looked into the PHP Zend Virtual Machine and found that the visibility mechanism is based on method flags as in Ruby.

**Python.** Python supports public and private visibility modifiers through name mangling [27]. By default, all attributes and methods are public, and the addition of a double underscore in front of their names marks them as private. Private attributes and methods are mangled including the name of the class, and call-sites are compiled making a syntactical difference between self-sends and object-sends. Each self-send using a private selector is compiled using the mangled name, restricting the visibility within the same class, but not its subclasses. Protected methods are marked by convention with a simple underscore but are not enforced.

Our solution uses a similar name-mangling process on protected methods. However, our name mangling is applied to all self and super-sends, only excluding messages implemented in public-only superclasses to avoid propagation (see Section 4.5). This allows developers to narrow recompilations.

**Javascript.** An ECMAScript proposal in stage 4 as of the writing of this paper proposes the inclusion of private fields to enforce encapsulation. Private fields are prefixed with a # and accessible only from inside the class defining it, encapsulating both properties and methods. Referring to # names outside the scope results in a syntax error at run time. In contrast, our solution proposes larger visibility while still supporting encapsulation and enforces visibility at compile-time using code rewrites.

**Java, C#, and C++.** In Java, C#, and C++ visibility modifiers are based on the type of the sender and receiver as well as on which package/assemblies they are defined in, they are not based on the identity of each instance. All three support the redefinition of virtual protected methods as public methods in subclasses. In our model, we support the same "opening" of protected methods in subclasses, since it means that from the redefinition of a protected method, object-sends would work as self-sends.

### 8.2 Object Encapsulation

Today, most statically-typed object-oriented languages such as Java, C++, and C# provide relatively good support for module encapsulation, and many proposals have been made for augmenting the static type systems of such languages so that they can also express object encapsulation [1, 2, 6, 7, 8, 20, 22, 24, 26]. We report here the work related to encapsulation espcially in dynamically-typed language.

In ConstrainedJava [19], the authors extend BeanShell, an extension of Java, to explore dynamic ownership in the context of a dynamically-typed language. Their model for dynamic ownership provides alias protection and encapsulation enforcement by maintaining a dynamic notion of object ownership at run time. It places restrictions on messages sent between objects based on their ownership. Their model further classifies message sends as *internal* (if the receiver of a message is this or is owned by this), or *visible* (sends to other visible objects). In addition, Dynamic Ownership recognizes two kinds of externally independent messages — pure messages that do not access the object state, and one-way messages that do not return results.

In the programming language MUST[37], methods can be private (visible only in the current class with here), public (visible everywhere), subclass-visible (callable with super), superclass-visible (callable with self), or both latest visibility. This encapsulation is based on a syntactic distinction: message-sends to any object, to self, to super, or to here are treated differently.

Schärli *et al.* proposed encapsulation policies as a way to constrain the interface of an object [30]. With Object-Oriented Encapsulation (OOE), two cases are distinguished: (1) an inheritance perspective where a class changes the way the superclass methods are bound from the subclass perspective and (2) an object perspective where the interface of an object itself is changed by associating encapsulation policies with object references. From the inheritance perspective, an encapsulation policy associated with a subclass changes how methods in the superclass are bound. Schärli *et al.* define three different rights to define the encapsulation of a method: the right to **{o}**verride, to r**{e}**-implement, and to **{c}**all a method. OOE semantics are also based on the syntactic distinction of three different messages: super-sends, self-sends, and object-sends. Only self-sends can be early bound.

Adding a protected method mechanism is closely tied to the implementation of encapsulation. The goal of OOE is broader than that of adding protected modifiers as presented in our model. What we want for a protected modifier is something between a **{co}** and an **{o}** policy.

In #PHARO, we make the same distinction between object-sends and self-sends: Object-sends can only access public methods, and self-sends access public or protected methods. This technique has also been chosen in [37, 12]. Finally, OOE required changes to the Virtual Machine implementation to include the new semantics, while our model does not require such changes and can be loaded as a library.

## 9 Conclusion

In dynamically-typed object-oriented languages, visibility modifiers are not as common or mature. In this paper, we presented PROTDYN, a self-send-based visibility model calculated at compile-time for dynamically-typed languages. Our model restricts protected methods activation to self/super-sends and makes them available only from an instance of the class defining them and their subclasses.

We present an implementation of PROTDYN in Pharo called #PHARO and a port to Python (See Appendix E). Our implementation is designed to be loadable as libraries and add negligible run-time costs. We show that the overhead introduced by our solution based on name-mangling is usually below 1 % and profits from common lookup optimizations such as global lookup caches and polymorphic inline caches. We show also that the introduction of protected methods would increase memory consumption of static code structures by 13 % in a worst-case scenario, but 2.1 % in a realistic scenario. Our solution is a viable approach to introduce a visibility modifiers in dynamic languages.

Iona Thomas, Vincent Aranega, Stéphane Ducasse, Guillermo Polito, and Pablo Tesone

## A  Definitions

Because the focus of this paper is on dynamically-typed languages, we briefly introduce some key vocabulary taken from the Smalltalk terminology.

**Definition 1. Message.** Messages are the key operations in object-oriented languages, also known in other languages as method invocations. A message is composed of a receiver, a selector, and zero or more arguments.

**Definition 2. Message receiver.** The message receiver is the object targeted by the message.

**Definition 3. Message selector.** The message selector is an identifier used to choose what method is executed. In dynamically-typed languages, a selector is used as a method signature. In statically-typed languages, the method signature contains a selector and also the types of its arguments and return value.

**Definition 4. Method lookup.** The method lookup is the process of searching for the right method to execute from a given message. In single-dispatched dynamically-typed object-oriented languages, the method lookup is a function on the receiver object and the selector. It looks up in the receiver's hierarchy the method whose signature is equal to the selector.

**Definition 5. Method activation.** A method activation is the execution of a method, triggered by a message-send. To activate a method, first, the method-lookup finds the corresponding method, then the method is executed on the message receiver and arguments.

**Definition 6. Current method receiver.** The current method receiver is the receiver object that led to the current method activation. It is usually denoted with special keywords or pseudo-variables named self or this.

**Definition 7. Message-send site.** A message send site is a location in the code where a message is sent, also known in other languages as a call site.

Now that the basic message-send terminology is set up, define the terminology related to method visibility. The following definitions are based on the previous fine-grained distinction between message, method activation, and message-send site.

**Definition 8. Method visibility.** A method is visible from a message-send site if the method lookup finds this method in the receiver's hierarchy. We say a send-site *can see* a method if the method is visible to the send-site.

**Definition 9. Visibility semantics.** The visibility semantics of a programming language are the rules that decide which methods are visible from which message-send sites.

**Definition 10. Visibility mechanism.** The visibility mechanism of a programming language is the technique used to guarantee that the language visibility semantics are not violated.

**Definition 11. self-send site or self-send.** A self-send site (self-send for short from now on) is a message-send site where we can syntactically identify the receiver as the current method receiver (*i.e.,* self or this). Unless they are specifically mentioned, we consider super-send sites as part of self-send sites.

**Definition 12. object-send site or object-send.** An object-send site (object-send for short from now on) is a message-send site where we cannot syntactically identify the receiver as the current method receiver (*i.e.,* it is not self or this).

**Definition 13. Protected and public methods.** A protected method is a method explicitly annotated by a developer as *protected*. A public method is a non-protected method.

## B  Object-Send and Self-Send Lookup Semantic by Example - Additional Explanation

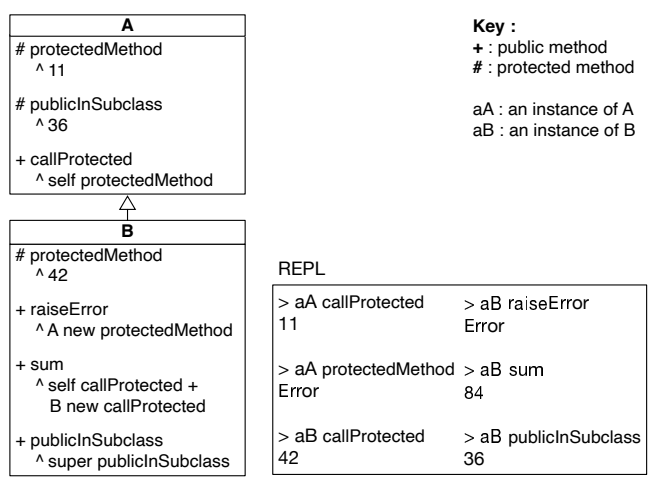

**Figure 13** Message sending is modified to distinguish between object-sends and self-sends: only self-sends can invoke protected methods which can also be overridden and taken into account by default method lookup.

**An object-send targetting a protected method produces an error.**
- Expression: `aA protectedMethod` – an instance of the class A is sent the message `protectedMethod`.
- Result: Error
- Explanation: The object-send `aA protectedMethod` does not find any public method with the selector `protectedMethod` and produces an error. Note that the method is not visible to this message send site even if at run time the receiver of `protectedMethod` is an instance of A because the receiver is not syntactically self nor super.

**A self-send targetting a protected method activates the method.**

- Expression: aA callProtected. – an instance of the class A is sent the message callProtected.
- Result: 11
- Explanation: The object-send aA callProtected finds the public method A»callProtected. From A»callProtected the self-send self protectedMethod finds and activates the protected method A»protectedMethod.

**Method lookup semantics are not modified by self-sends.**

- Expression: aB callProtected. – an instance of the class B is sent the message callProtected.
- Result: 42
- Explanation: The object-send aB callProtected finds the public method A»callProtected and activates it on aB. From A»callProtected the self-send self protectedMethod finds and activates the overridden protected method B»protectedMethod because aB the receiver is an instance of B.

**Protected method visibility is not lexically bound.**

- Expression: aB raiseError. – an instance of the class B is sent the message raiseError.
- Result: Error
- Explanation: The object-send aB raiseError finds the public method B»raiseError and activates it on aB. From B»raiseError the object-send A new protectedMethod looks for a public method with selector protectedMethod, finds none, and produces an error. Notice again that protected methods are not visible to this message send site even if the lexical scope contains a definition of protectedMethod.

**Any message-send targetting a public method activates the method.**

- Expression: aB sum. – an instance of the class B is sent the message sum.
- Result: 84
- Explanation: The object-send aB sum finds the public method B»sum and activates it on aB. From B»sum (a) the self-send self callProtected finds and activates the superclass' public method A»callProtected and (b) the object-send B new callProtected finds and activates the superclass' public method A»callProtected.

**Increasing visibility in subclasses.**

- Expression: aB publicInSubclass. – an instance of the class B is sent the message publicInSubclass.
- Result: 36
- Explanation: The object-send aB publicInSubclass finds the public method B»publicInSubclass and activates it on aB. From B»publicInSubclass the self-send super publicInSubclass finds and activates the superclass' protected method A»publicInSubclass. This example shows that subclasses can redefine and increase of visibility of protected methods in subclasses. As explained in the next section, restricting visibility is not allowed by construction.

## C SMALLTALKLITE

We present SMALLTALKLITE, a Smalltalk-like dynamic language featuring single inheritance, message-passing, field access and update, and **self** and **super** sends. SMALLTALKLITE is similar to CLASSICJAVA, but removes interfaces and static types. Fields are private in SMALLTALKLITE, so only local or inherited fields may be accessed.

We base our approach on the object model used by Flatt *et al.* [16] to give a semantic for mixins for Java-like languages. We modify their CLASSICJAVA model to develop SMALLTALKLITE, a simple calculus that captures the key features of Smalltalk-like dynamic languages.

### C.1 SMALLTALKLITE Reduction Semantics

The syntax of SMALLTALKLITE is shown in Figure 15. SMALLTALKLITE is similar to CLASSICJAVA while eliding the features related to static typing. We similarly ignore features that are not relevant to a discussion of traits, such as reflection or class-side methods.

To simplify the reduction semantics of SMALLTALKLITE, we adopt an approach similar to that used by Flatt *et al.* [16], namely we annotate field accesses and **super** sends with additional static information that is needed at "run-time". This extended redex syntax is shown in Figure 14. The figure also specifies the evaluation contexts for the extended redex syntax in Felleisen and Hieb's notation [15].

Predicates and relations used by the semantic reductions are listed in Figure 17. (The predicates CLASSESONCE($P$) *etc* are assumed to be preconditions for valid programs, and are not otherwise explicitly mentioned in the reduction rules.)

$P \vdash \langle \epsilon, \mathscr{S} \rangle \hookrightarrow \langle \epsilon', \mathscr{S}' \rangle$ means that we reduce an expression (redex) $\epsilon$ in the context of a (static) program $P$ and a (dynamic) store of objects $\mathscr{S}$ to a new expression $\epsilon'$ and (possibly) updated store $\mathscr{S}'$. A redex $\epsilon$ is essentially an expression $e$ in which field names are decorated with their object contexts, *i.e.*, $f$ is translated to $o.f$, and **super** sends are decorated with their object and class contexts. Redexes and their subexpressions reduce to a value, which is either an object identifier or nil. Subexpressions may be evaluated within an expression context $E$.

The store consists of a set of mappings from object identifiers $oid \in \text{dom}(\mathscr{S})$ to tuples $\langle c, \{f \mapsto v\} \rangle$ representing the class $c$ of an object and the set of its field values. The initial value of the store is $\mathscr{S} = \{\}$.

Translation from the main expression to an initial redex is specified by the $o[\![e]\!]_c$ function (see Figure 16). This binds fields to their enclosing object context and binds **self** to the *oid* of the receiver. The initial object context for a program is nil. (*i.e.*, there are no global fields accessible to the main expression). So if $e$ is the main expression associated with a program $P$, then $\text{nil}[\![e]\!]_{\text{Object}}$ is the initial redex.

The reductions are summarised in Figure 18.

**new** $c$ [*new*] reduces to a fresh *oid*, bound in the store to an object whose class is $c$ and whose fields are all nil. A (local) field access [*get*] reduces to the value of the field. Note that it is syntactically impossible to access a field of another object. The

$$
\begin{aligned}
\epsilon \; &= \; v \mid \textbf{new}\; c \mid x \mid \textbf{self} \mid \epsilon.f \mid \epsilon.f{=}\epsilon \\
&\mid \; \epsilon.m(\epsilon^*) \mid \textbf{super}\langle o,c\rangle.m(\epsilon^*) \mid \textbf{let}\; x{=}\epsilon \; \textbf{in}\; \epsilon \\
E \; &= \; [\;] \mid o.f{=}E \mid E.m(\epsilon^*) \mid o.m(v^*\; E\; \epsilon^*) \\
&\mid \; \textbf{super}\langle o,c\rangle.m(v^*\; E\; \epsilon^*) \mid \textbf{let}\; x{=}E \; \textbf{in}\; \epsilon \\
v,o \; &= \; \textsf{nil} \mid \textit{oid}
\end{aligned}
$$

**Figure 14** Redex syntax

$$
\begin{aligned}
P \; &= \; \textit{defn}^*\, e \\
\textit{defn} \; &= \; \textbf{class}\; c\; \textbf{extends}\; c\; \{\; f^*\textit{meth}^*\; \} \\
e \; &= \; \textbf{new}\; c \mid x \mid \textbf{self} \mid \textsf{nil} \\
&\mid \; f \mid f{=}e \mid e.m(e^*) \\
&\mid \; \textbf{super}.m(e^*) \mid \textbf{let}\; x{=}e\; \textbf{in}\; e \\
\textit{meth} \; &= \; m(x^*)\; \{\; e\; \} \\
c \; &= \; \text{a class name} \mid \textsf{Object} \\
f \; &= \; \text{a field name} \\
m \; &= \; \text{a method name} \\
x \; &= \; \text{a variable name}
\end{aligned}
$$

**Figure 15** SMALLTALKLITE syntax

$$
\begin{aligned}
o[\![\textbf{new}\; c']\!]_c \; &= \; \textbf{new}\; c' \\
o[\![x]\!]_c \; &= \; x \\
o[\![\textbf{self}]\!]_c \; &= \; o \\
o[\![\textsf{nil}]\!]_c \; &= \; \textsf{nil} \\
o[\![f\,]\!]_c \; &= \; o.f \\
o[\![f{=}e]\!]_c \; &= \; o.f{=}o[\![e]\!]_c \\
o[\![e.m(e_i^*)]\!]_c \; &= \; o[\![e]\!]_c.m(o[\![e_i]\!]_c^*) \\
o[\![\textbf{super}.m(e_i^*)]\!]_c \; &= \; \textbf{super}\langle o,c\rangle.m(o[\![e_i]\!]_c^*) \\
o[\![\textbf{let}\; x{=}e\; \textbf{in}\; e']\!]_c \; &= \; \textbf{let}\; x{=}o[\![e]\!]_c\; \textbf{in}\; o[\![e']\!]_c
\end{aligned}
$$

**Figure 16** Translating expressions to redexes

redex notation $o.f$ is only generated in the context of the object $o$. Field update [*set*] simply updates the corresponding binding of the field in the store.

When we send a message [*send*], we must look up the corresponding method body $e$, starting from the class $c$ of the receiver $o$. The method body is then evaluated in the context of the receiver $o$, binding **self** to the receiver's *oid*. Formal parameters to the method are substituted by the actual arguments (see Figure 19). We also pass in the actual class in which the method is found, so that **super** sends have the right context to start their method lookup.

| | |
|---|---|
| $\prec_P$ | Direct subclass |
| | $c \prec_P c' \iff$ **class** $c$ **extends** $c' \cdots \{\cdots\} \in P$ |
| $\leq_P$ | Indirect subclass |
| | $c \leq_P c' \equiv$ transitive, reflexive closure of $\prec_P$ |
| $\in_P$ | Field defined in class |
| | $f \in_P c \iff$ **class** $\cdots \{\cdots f \cdots\} \in P$ |
| $\in_P$ | Method defined in class |
| | $\langle m, x^*, e \rangle \in_P c \iff$ **class** $\cdots \{\cdots m(x^*)\{e\}\cdots\} \in P$ |
| $\in_P^*$ | Field defined in $c$ |
| | $f \in_P^* c \iff \exists c', c \leq_P c', f \in_P c'$ |
| $\in_P^*$ | Method lookup starting from $c$ |
| | $\langle c, m, x^*, e \rangle \in_P^* c' \iff c' = min\{c'' \mid \langle m, x^*, e \rangle \in_P c'', c \leq_P c''\}$ |
| ClassesOnce($P$) | Each class name is declared only once |
| | $\forall c, c',$ **class** $c \cdots$ **class** $c' \cdots$ is in $P \Rightarrow c \neq c'$ |
| FieldOncePerClass($P$) | Field names are unique within a class declaration |
| | $\forall f, f',$ **class** $c \cdots \{\cdots f \cdots f' \cdots\}$ is in $P \Rightarrow f \neq f'$ |
| FieldsUniquelyDefined($P$) | Fields cannot be overridden |
| | $f \in_P c, c \leq_P c' \implies f \notin_P c'$ |
| MethodOncePerClass($P$) | Method names are unique within a class declaration |
| | $\forall m, m',$ **class** $c \cdots \{\cdots m(\cdots)\{\cdots\} \cdots m'(\cdots)\{\cdots\} \cdots\}$ is in $P \Rightarrow m \neq m'$ |
| CompleteClasses($P$) | Classes that are extended are defined |
| | range($\prec_P$) $\subseteq$ dom($\prec_P$) $\cup$ {Object} |
| WellFoundedClasses($P$) | Class hierarchy is an order |
| | $\leq_P$ is antisymmetric |
| ClassMethodsOK($P$) | Method overriding preserves arity |
| | $\forall m, m', \langle m, x_1 \cdots x_j, e \rangle \in_P c, \langle m, x'_1 \cdots x'_k, e' \rangle \in_P c', c \leq_P c' \implies j = k$ |

**Figure 17** Relations and predicates for SmalltalkLite

**super** sends [*super*] are similar to regular message sends, except that the method lookup must start in the superclass of the class of the method in which the **super** send was declared. When we reduce the **super** send, we must take care to pass on the class $c''$ of the method in which the **super** method was found, since that method may make further **super** sends. **let in** expressions [*let*] simply represent local variable bindings.

Errors occur if an expression gets "stuck" and does not reduce to an *oid* or nil. This occurs if a non-existent variable, field, or method is referenced (for example, when sending a message to nil). For the purpose of this paper, we are not concerned with errors, so we do not introduce any special rules for these cases.

## D ProtectedLite: ProtDyn Semantics

To specify the semantics of self-sends versus object-sends, we define ProtectedLite, an extension of SmalltalkLite [4]. SmalltalkLite is a dynamic language calculus featuring single inheritance, message-passing, field access and updates, and self/super sends. The syntax used in the calculus is presented in Figure 20. SmalltalkLite is heavily inspired by ClassicJava defined by Flatt *et al.* [16]. We do not consider it to be a contribution of this article. We repeated the full description of SmalltalkLite in the appendix to help the reader.

Iona Thomas, Vincent Aranega, Stéphane Ducasse, Guillermo Polito, and Pablo Tesone

$$P \;\vdash\; \langle E[\mathbf{new}\; c], \mathscr{S}\rangle \hookrightarrow \langle E[oid], \mathscr{S}[oid \mapsto \langle c, \{f \mapsto \mathrm{nil} \mid \forall f, f \in_P^* c\}\rangle]\rangle \qquad [new]$$
where $oid \notin \mathrm{dom}(\mathscr{S})$

$$P \;\vdash\; \langle E[o.f], \mathscr{S}\rangle \hookrightarrow \langle E[v], \mathscr{S}\rangle \qquad [get]$$
where $\mathscr{S}(o) = \langle c, \mathscr{F}\rangle$ and $\mathscr{F}(f) = v$

$$P \;\vdash\; \langle E[o.f=v], \mathscr{S}\rangle \hookrightarrow \langle E[v], \mathscr{S}[o \mapsto \langle c, \mathscr{F}[f \mapsto v]\rangle]\rangle \qquad [set]$$
where $\mathscr{S}(o) = \langle c, \mathscr{F}\rangle$

$$P \;\vdash\; \langle E[o.m(v^*)], \mathscr{S}\rangle \hookrightarrow \langle E[o[\![e[v^*/x^*]]\!]_{c'}], \mathscr{S}\rangle \qquad [send]$$
where $\mathscr{S}(o) = \langle c, \mathscr{F}\rangle$ and $\langle c, m, x^*, e\rangle \in_P^* c'$

$$P \;\vdash\; \langle E[\mathbf{super}\langle o, c\rangle.m(v^*)], \mathscr{S}\rangle \hookrightarrow \langle E[o[\![e[v^*/x^*]]\!]_{c''}], \mathscr{S}\rangle \qquad [super]$$
where $c \prec_P c'$ and $\langle c', m, x^*, e\rangle \in_P^* c''$ and $c' \leq_P c''$

$$P \;\vdash\; \langle E[\mathbf{let}\; x=v\; \mathbf{in}\; \epsilon], \mathscr{S}\rangle \hookrightarrow \langle E[\epsilon[v/x]], \mathscr{S}\rangle \qquad [let]$$

**Figure 18** Reductions for SMALLTALKLITE

$$
\begin{aligned}
\mathbf{new}\; c\; [v/x] &= \mathbf{new}\; c \\
x\; [v/x] &= v \\
x'\; [v/x] &= x' \\
\mathbf{self}\; [v/x] &= \mathbf{self} \\
\mathrm{nil}\; [v/x] &= \mathrm{nil} \\
f\; [v/x] &= f \\
f{=}e\; [v/x] &= f{=}e[v/x] \\
e.m(e_i^*)\; [v/x] &= e[v/x].m(e_i^*[v/x]) \\
\mathbf{super}.m(e_i^*)\; [v/x] &= \mathbf{super}.m(e_i^*[v/x]) \\
\mathbf{let}\; x{=}e\; \mathbf{in}\; e'\; [v/x] &= \mathbf{let}\; x{=}e[v/x]\; \mathbf{in}\; e' \\
\mathbf{let}\; x'{=}e\; \mathbf{in}\; e'\; [v/x] &= \mathbf{let}\; x'{=}e[v/x]\; \mathbf{in}\; e'[v/x]
\end{aligned}
$$

**Figure 19** Variable substitution

$$
\begin{array}{rcl}
P &=& defn^* e \\
defn &=& \mathbf{class}\; c\; \mathbf{extends}\; c'\; \{\; f^* meth^* protectmeth^*\; \} \\
e &=& \mathbf{new}\; c \mid x \mid \mathbf{self} \mid \mathrm{nil} \\
  &\mid& f \mid (f{=}e) \mid e.m(e^*) \\
  &\mid& \mathbf{super}.m(e^*) \mid \mathbf{let}\; x{=}e\; \mathbf{in}\; e
\end{array}
\qquad
\begin{array}{rcl}
meth &=& m(x^*)\;\{\;e\;\} \\
protectmeth &=& \#m(x^*)\;\{\;e\;\} \\
c &=& \text{a class name} \mid \mathrm{Object} \\
f &=& \text{a field name} \\
m &=& \text{a method name} \\
x &=& \text{a variable name}
\end{array}
$$

**Figure 20** Protected Pharo syntax.

Each class in PROTECTEDLITE has a list of protected methods after the public ones (see Figure 20). The MethodOncePerClass predicate presented in Figure 21 specifies that two methods shall not have the same name, even if they have different modifiers.

The OverridingPublicMethod and OverridingProtectedMethod predicates guarantee that public methods can be only overridden by public methods and that protected methods can be overridden by public or protected methods. These predicates guarantee per construction that is not possible to restrict the visibility of a public method.

| | |
|---|---|
| MethodOncePerClass($P$) | Method names are unique within a class declaration |
| | $\forall m, m', \#m, \#m'$ |
| | **class** $c \cdots \{\cdots m(\cdots)\{\cdots\} \cdots m'(\cdots)\{\cdots\} \cdots$ |
| | $\#m(\cdots)\{\cdots\} \cdots \#m'(\cdots)\{\cdots\}\}$ is in $P$ |
| | $\Rightarrow Set(m, m', \#m, \#m')size = 4$ |
| OverridingProtectedMethod($P$) | Protected methods can be overridden by public |
| | or protected methods |
| | $\forall m, \langle m, x^*, e \rangle \widehat{\in_P} c, \langle m, x^*, e \rangle \in_P c', c \leq_P c'$ |
| | or $\forall m, \langle m, x^*, e \rangle \widehat{\in_P} c, \langle m, x^*, e \rangle \widehat{\in_P} c', c \leq_P c'$ |
| OverridingPublicMethod($P$) | Public methods can be overridden only by public methods |
| | $\forall m, \langle m, x^*, e \rangle \in_P c, \langle m, x^*, e \rangle \in_P c', c \leq_P c'$ |
| $\in_P$ | Public method defined in class |
| | $\langle m, x^*, e \rangle \in_P c \iff \textbf{class} \cdots \{\cdots m(x^*)\{e\}\cdots\} \in P$ |
| $\widehat{\in_P}$ | Protected method defined in class |
| | $\langle m, x^*, e \rangle \widehat{\in_P} c \iff \textbf{class} \cdots \{\cdots \#m(x^*)\{e\}\cdots\} \in P$ |
| $\in_P^*$ | Public method lookup starting from $c$ |
| | $\langle c, m, x^*, e \rangle \in_P^* c' \iff c' = min\{c'' \mid \langle m, x^*, e \rangle \in_P c'', c \leq_P c''\}$ |
| $\widehat{\in_P^*}$ | Protected method lookup starting from $c$ |
| | $\langle c, m, x^*, e \rangle \widehat{\in_P^*} c' \iff c' = min\{c'' \mid \langle \#m, x^*, e \rangle \widehat{\in_P} c'', c \leq_P c''\}$ |

**Figure 21** Relations and predicates for ProtectedLite.

$P \vdash \langle E[o.m(v^*)], \mathscr{S} \rangle \hookrightarrow \langle E[o[\![e[v^*/x^*]]\!]_{c'}], \mathscr{S} \rangle$     [*object-send*]
where $\mathscr{S}[o] = \langle c, \mathscr{F} \rangle$ and $\langle c, m, x^*, e \rangle \in_P^* c'$

$P \vdash \langle E[\textbf{self}\langle o, c \rangle.m(v^*)], \mathscr{S} \rangle \hookrightarrow \langle E[o[\![e[v^*/x^*]]\!]_{c'}], \mathscr{S} \rangle$     [*self-send*]
where $\mathscr{S}[o] = \langle c, \mathscr{F} \rangle$ and $\langle c, m, x^*, e \rangle \widehat{\in_P^*} c'$ or $\langle c, m, x^*, e \rangle \in_P^* c'$

$P \vdash \langle E[\textbf{super}\langle o, c \rangle.m(v^*)], \mathscr{S} \rangle \hookrightarrow \langle E[o[\![e[v^*/x^*]]\!]_{c''}], \mathscr{S} \rangle$     [*super-send*]
where $c \prec_P c'$, and $\langle c', m, x^*, e \rangle \widehat{\in_P^*} c''$ or $\langle c', m, x^*, e \rangle \in_P^* c''$, and $c' \leq_P c''$

**Figure 22** Message passing reductions for ProtectedLite.

The next four predicates express the lookup mechanism used in ProtectedLite. Predicates $\in_P^*$ and $\widehat{\in_P^*}$ express the lookup of public and protected methods respectively. The lookup mechanism finds the closest superclass ($c'$) of the receiver class ($c$) that contains a method with the given selector ($m'$). The main difference between both mechanisms is that the public lookup only searches in public methods and the protected lookup only in protected ones.

The public and protected lookup mechanisms are then used in Figure 22 to define the self-send, super-send, and object-send. When a message is sent to an object (without using self or super), we look up the method body, starting from the class of the object $c$ and only look at public methods. For both self-send and super-send, we use both lookups. Super-sends use the same mechanism as self-sends, but the lookup starts from the superclass ($c'$) of the class defining the method.

**Selector mangling for self-sends sites**    As we saw in Section 4.4, during compilation, all self-sends are mangled. Figure 23 formalizes this transformation. The expression

$$
\begin{aligned}
\mathbf{new}\ c\ [m'/m]_{self} &= \mathbf{new}\ c & \text{instance creation} \\
\mathbf{let}\ x = e\ \mathbf{in}\ e'\ [m'/m]_{self} &= \mathbf{let}\ x = e[m'/m]_{self}\ \mathbf{in}\ e'[m'/m]_{self} \\
x\ [m'/m]_{self} &= x & \text{variable access} \\
f\ [m'/m]_{self} &= f & \text{field access} \\
(f = e)\ [m'/m]_{self} &= (f = e[m'/m]_{self}) & \text{field assignment} \\
\mathbf{nil}\ [m'/m]_{self} &= \mathbf{nil} \\
\\
\mathbf{self}.m(e_i^*)\ [m'/m]_{self} &= \mathbf{self}.m'(e_i^*[m'/m]_{self}) & \text{self/super sends} \\
\mathbf{self}.n(e_i^*)\ [m'/m]_{self} &= \mathbf{self}.n(e_i^*[m'/m]_{self}),\ \text{if}\ n \neq m \\
\mathbf{super}.m(e_i^*)\ [m'/m]_{self} &= \mathbf{super}.m'(e_i^*[m'/m]_{self}) & \text{can call protected methods} \\
\mathbf{super}.n(e_i^*)\ [m'/m]_{self} &= \mathbf{super}.n(e_i^*[m'/m]_{self}),\ \text{if}\ n \neq m \\
e.m(e_i^*)\ [m'/m]_{self} &= e[m'/m]_{self}.m(e_i^*[m'/m]_{self}) & \text{object-sends only} \\
& & \text{call public methods}
\end{aligned}
$$

■ **Figure 23** Self and super-send call site renaming (also called selector mangling). Method names of object-sends are not renamed to protected method name mangling, while self and super sends are.

$[m'/m]_{self}$ renames all the self-sends to $m$ to $m'$, where $m'$ is the mangled selector obtained from the $\rho_{[]}$ hiding function. The scope of this hiding function is global.

## E  Applicability Outside Pharo: PROTDYN Python Port

Python does not provide the same facilities as Pharo regarding the modification of the internal compiler directly from the language itself. We use Python decorators to mark methods as @protected, as illustrated in Listing 2 with the protectedMethod method, and two hooks of Python's Meta-Object-Protocol to detect class creation.

■ **Listing 2** Excerpt of the example in Figure 2 reimplemented in Python using a @protected decorator as protected modifier.

```python
from protected import protected
# ...
class B(A):
  @protected
  def protectedMethod(self):
    return 42

  def raiseError(self):
    return A().protectedMethod()

  def sum(self):
    return self.callProtected() + B().callProtected()

  def publicInSubclass(self):
    return super().publicInSubclass()
```

The protected decorator registers a callback method — using the __set_name__ hook[1]— that is automatically called when the owning class object is created. When this callback is triggered, the class is recompiled as well as its superclasses. The recompilation of a class is made of two major steps: the modification of all self-send sites and the installation of the non-protected methods under both the original selector and the mangled selector in the class.

Finally, we register a callback on each class at the top of the inheritance tree to customize subclass creation using the __subclass_init__ hook.[2] The customization of the subclass creation triggers the recompilation process for each new subclass created, even if those subclasses do not define protected methods. *Note:* the registration of this dedicated callback for each top class ensures that the subclasses customization process supports multiple inheritance.

**Limitations:** The recompilation process triggered by the __set_name__ and __subclass_init__ hooks only deals with methods that are declared in their classes. Our current implementation does not deal with dynamically added methods (monkey patching) on a class that defines protected methods in its inheritance tree. To address this issue, the implementation can rely on the change of the normal Python metaclass of each class at the top of the inheritance tree for a special Python metaclass that registers a callback on attribute injection in the class.

## F  Numerical Results for Benchmarks

The following table is showing the results of the three runtime performance experiments described in Section 5.3

■ **Table 3**  Average run times and standard interval with confidence of 95 % for the different benchmarks: without using the protected modifier library, with the protected modifier library loaded and with the protected modifier library used

| Experiment | Benchmarks | Without protected | Loaded | Used |
|---|---|---|---|---|
| 1 | Microdown | 1880.8ms ± 0.4 ms | 1880.3ms ± 0.3ms | 1878.9ms ± 0.4ms |
|  | Delta Blue | 8286ms ± 5ms | 8272ms ± 5ms | 8197ms ± 6ms |
|  | Richards | 3519ms ± 4ms | 3522ms ± 4ms | 3522ms ± 4ms |
|  | Compiler | 4225ms ± 2ms | 4211ms ± 2ms | 4243ms ± 2ms |
| 2 | Microdown | 788.2ms ± 0.7ms | 797.1ms ± 0.6ms | 797.2ms ± 0.6ms |
|  | Delta Blue | 4197ms ± 2ms | 4219ms ± 5ms | 4180ms ± 2ms |
|  | Richards | 2148ms ± 2ms | 2134ms ± 2ms | 2111ms ± 2ms |
|  | Compiler | 2891ms ± 2ms | 2889ms ± 2ms | 2921ms ± 2ms |
| 4 | Microdown | 441.4ms ± 0.5ms | 440.8ms ± 0.4ms | 439.9ms ± 0.4ms |
|  | Delta Blue | 1275ms ± 3ms | 1281ms ± 2ms | 1283ms ± 3ms |
|  | Richards | 448.2ms ± 0.7ms | 461.1ms ± 0.8ms | 458.9ms ± 0.8ms |
|  | Compiler | 906.6ms ± 1.6ms | 914.2ms ± 1.6ms | 919.7ms ± 1.6ms |

---

[1] https://docs.python.org/3/reference/datamodel.html (Accessed on 2023/06/01)
[2] https://docs.python.org/3/reference/datamodel.html#object.__init_subclass__ (Accessed on 2023/06/01)

A VM-Agnostic and Backwards Compatible Protected Modifier f. Dynamically-Typed Lang.

## About the authors

**Iona Thomas** iona.thomas@inria.fr
https://orcid.org/0000-0001-8490-3802

**Vincent Aranega** vincent.aranega@inria.fr
https://orcid.org/0000-0003-4465-1289

**Stéphane Ducasse** stephane.ducasse@inria.fr
https://orcid.org/0000-0001-6070-6599

**Guillermo Polito** guillermo.polito@inria.fr
https://orcid.org/0000-0003-0813-8584

**Pablo Tesone** pablo.tesone@inria.fr
https://orcid.org/0000-0002-5615-6691